\documentclass[%
reprint,
superscriptaddress,
amsmath,amssymb,
aps,prl
]{revtex4-1}

\usepackage{graphicx}
\usepackage{dcolumn}
\usepackage{bm}
\usepackage{hyperref}
\usepackage{float}
\usepackage{cases}
\usepackage{cleveref}
\usepackage{wasysym}
\usepackage[makeroom]{cancel}
\usepackage{scalerel}
\usepackage{lipsum}
\usepackage{svg}
\usepackage{soul}
\usepackage{xcolor}

\hypersetup{
	colorlinks,
	linkcolor={red},
	citecolor={blue},
	urlcolor={black}
}

\begin{document}

	\preprint{APS/123-QED}
	\sloppy
	\allowdisplaybreaks
	\title{Self-assembly Through Programmable Folding}
	\author{Angus McMullen}
	\affiliation{Center for Soft Matter Research, New York University, New York, NY, USA}
	\author{Maitane Muñoz Basagoiti}
    \thanks{A.M. and M.M.B. contributed equally to this work.}
	\affiliation{Gulliver UMR CNRS 7083, ESPCI Paris, Université PSL, 75005 Paris, France}
	\author{Zorana Zeravcic}
	\affiliation{Gulliver UMR CNRS 7083, ESPCI Paris, Université PSL, 75005 Paris, France}
	\author{Jasna Brujic}
	\affiliation{Center for Soft Matter Research, New York University, New York, NY, USA}
	\email{jb2929@nyu.edu}

	\date{\today}

	\maketitle

\textbf{At the cutting edge of materials science, matter is designed to self-organize into structures that perform a wide range of functions \cite{talapin2020functional}. The past two decades have witnessed major innovations in the versatility of building blocks, ranging from DNA on the nanoscale to handshaking materials on the macroscale \cite{rothemund2006folding, douglas2009self, winfree1998design, nykypanchuk2008dna,nykypanchuk2008dna, rogers2015programming,du2021programming, niu2019magnetic}. Like a jigsaw puzzle, one can reliably self-assemble arbitrary structures if all the pieces are distinct \cite{ke2012three,Halverson2013,zeravcic2014size,Ong:2017jx}, but systems with fewer flavors of building blocks have so far been limited to the assembly of exotic crystals \cite{he2020colloidal,Macfarlane2011,lin2017clathrate}. Inspired by Nature’s strategy of folding biopolymers into specific RNA and protein structures,  here we introduce a minimal model system of colloidal polymers with programmable DNA interactions that guide their downhill folding into two-dimensional geometries. Combining experiments, simulations, and theory, we show that designing the order in which interactions are switched on directs folding into unique geometries called foldamers. The simplest alternating sequences ($ABAB$...) of up to 13 droplets yield eleven foldamers, while designing the sequence and adding an extra flavor uniquely encodes more than half of the 619 possible geometries. These foldamers can further interact to make complex supracolloidal architectures, seeding a next generation of bio-inspired materials. Our results are independent of the dynamics and therefore apply to polymeric materials with hierarchical interactions on all length scales, from organic molecules all the way to Rubik’s snakes.
}

\begin{figure*}[t!]
    \centering
   \includegraphics[width = \textwidth]{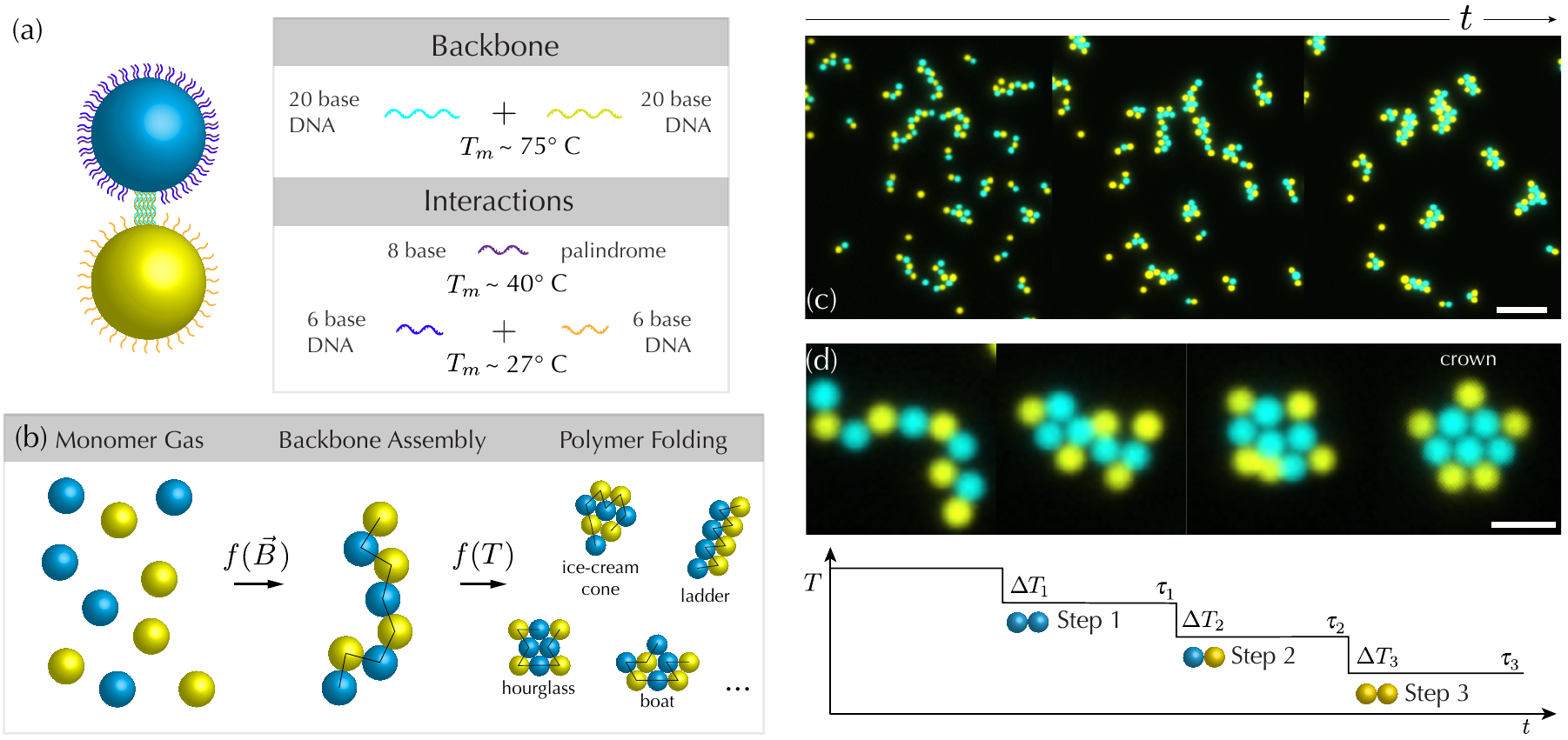}
   \caption{(a) Two flavors of droplets, $A$ (blue) and $B$ (yellow) are functionalized with complementary backbone strands of DNA to make alternating chains. They also carry weaker DNA interactions that mediate folding. (b) An emulsion first assembles into colloidomers using a magnetic field, after which a temperature protocol triggers folding into diverse geometries. (c) Fluorescent images show colloidomers of different lengths that undergo folding over time. Scale bar is 20\,$\mu$m. (d) A temperature protocol gives rise to stepwise folding of a decamer chain into the crown foldamer. Scale bar is 5\,$\mu$m.
}
   \label{fig:introduction}
\end{figure*}
	\vspace*{5mm}
Self-assembly of materials currently requires a toolbox of building blocks with a given shape and a multitude of interaction flavors and strengths to ensure a unique product. The underlying design principles either employ thermodynamics in search of a crystalline global free energy minimum \cite{wang2017colloidal, he2020colloidal, lin2017clathrate, rogers2015programming, nykypanchuk2008dna, VanAnders2014, lu2015superlattices, dijkstra2021predictive}, or prescribe sufficient specificity such that the building blocks can assemble one \cite{Zion2017, Zhang2018, ke2012three, ke2014dna,Ong:2017jx} or several target structures \cite{Murugan2014}. Despite these advances, achieving self-assembly of an arbitrary structure with high yield using a limited palette of flavors remains a key challenge.

In search of an alternative approach for the design of functional materials, we turn to the biological concept of self-assembly via folding. At the molecular level, RNA and proteins robustly fold into well-defined structures starting from evolution-selected sequences of only 4 nucleotides or 20 amino acids, respectively. On the cellular level, folded proteins self-organize into higher level complexes, such as motors or microtubules, whose functionality depends on the shape of the constituents \cite{Levy2006}.

Adopting the strategy of folding to materials science, here we demonstrate that colloidal droplet chains, i.e., colloidomers, endowed with few flavors can be programmed to fold into a selection of foldamers with near-perfect yield. The geometric shape and specificity of foldamer interactions can then be tuned to give rise to their self-assembly into supracolloidal architectures, such as dimers, micelles, tubules, and mosaic tilings. This novel toolbox enables the encoding of large-scale design into sequences of short linear polymers, placing folding at the forefront of materials self-assembly.

\vspace{1mm}
\noindent {\bf A folding model system}
\vspace{1mm}

Our system consists of two flavors of colloidal droplets, labeled blue ($A$) and yellow ($B$), functionalized with complementary DNA strands (Methods). These droplets irreversibly bind with valence two to form the backbone of an alternating colloidomer \cite{McMullen2018, McMullen2021}, depicted in Fig.\,\ref{fig:introduction} (a,b). The droplets are dispersed in an aqueous ferrofluid and we apply an intermittent magnetic field to accelerate the chaining process. These chains are thermal and freely jointed because DNA diffuses on the surface even after the droplets are bound.

To mediate folding, each droplet flavor is additionally functionalized with DNA strands that act as weaker secondary interactions. If they are all simultaneously switched on, one obtains a mixture of folded geometries as the final product \cite{Trubiano2021}.
The number of possible geometries is singular for chains shorter than hexamers, but then grows exponentially with chain length. For example, an octamer can fold into nine distinct geometries, four of which are shown in Fig.\,\ref{fig:introduction} (b). By choosing strands with distinct binding energies and therefore different melting temperatures \cite{Wang2015, lowensohn2019linker, gehrels2018using} (Methods), we establish a hierarchy of bonds that are switched on as the temperature is lowered, shown in Fig.\,\ref{fig:introduction} (c,d). Because the melting transition is sharp, working a few degrees below it ensures irreversible bond formation and downhill folding. For example, the decamer chain in (d) folds into the crown in a stepwise manner. First, the blue-blue palindrome interaction forms a pentamer core at high temperature, followed by the sequential locking in of yellow-blue and yellow-yellow bonds at progressively lower temperatures.
\begin{figure*}
  \centering
  \includegraphics{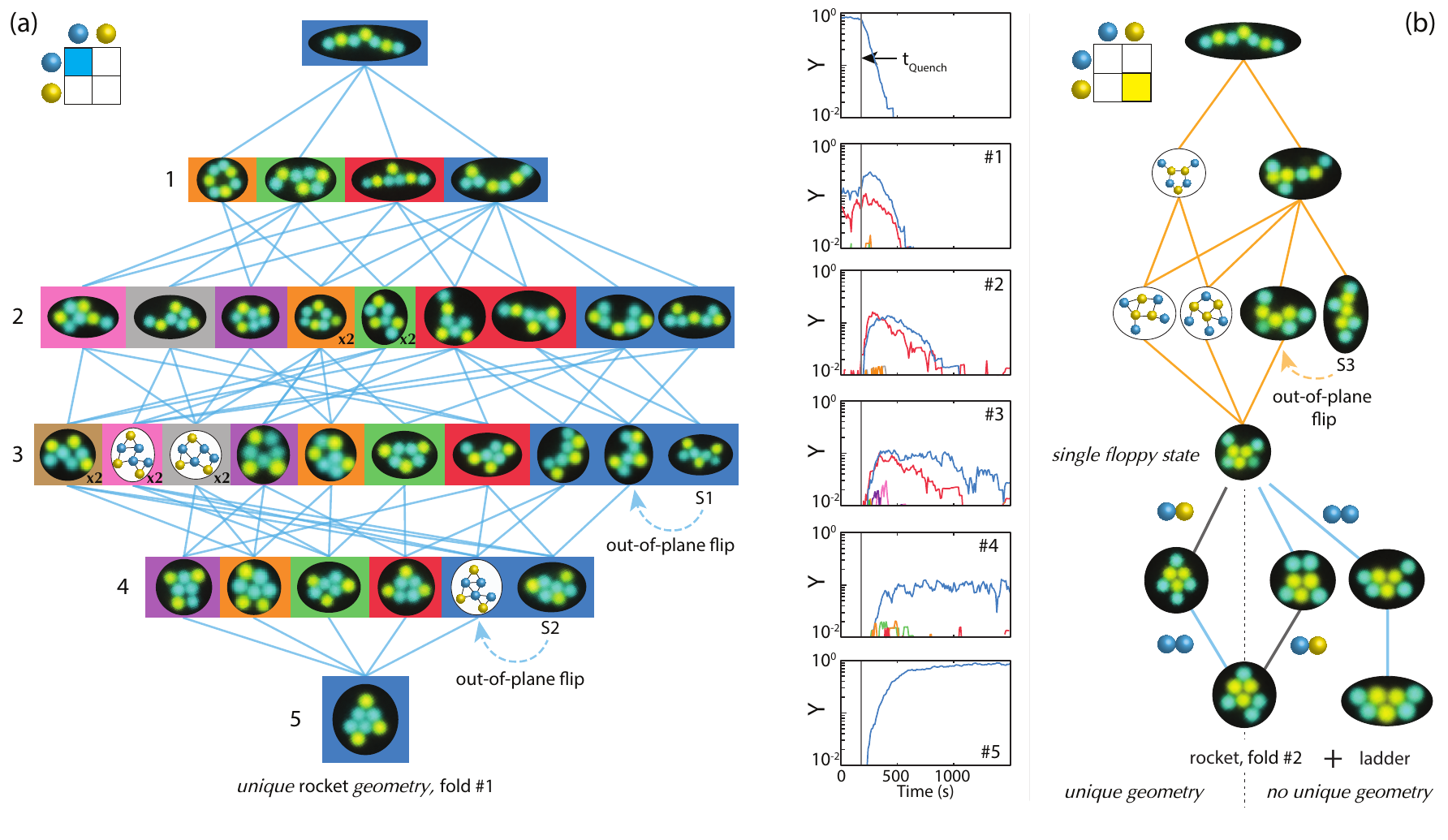} 
  \caption{
  (a) All folding pathways of a four blue{,} three yellow droplets heptamer result in a rocket foldamer when only the blue-blue interaction is turned on. Experimental images of states are superimposed with the theoretical tree, in order of frequency, to show the diversity of observed pathways. The number of secondary bonds acquired is shown at each level of the tree. The plots on the right show the time evolution after the temperature quench $t_{quench}$ of the yield of each color-coordinated state. (b) When the yellow-yellow interaction is switched on first, the same polymer folds into a single floppy state. Further interactions  fold it into a rocket with a different fold, but reversing the order of interactions leads to a mixture of the rocket and the ladder.}
  \label{fig:example}
\end{figure*}

\vspace{-2mm}
\noindent {\bf Design of the folding landscape}
\vspace{1mm}

Along the folding process, each new bond that forms causes the chain to adopt a different configuration. Those configurations that have the same contact matrix, ignoring chirality, are here defined to belong to a given state. All possible states between the linear chain and the final geometries map out an energy landscape that can be represented in a tree form. In the folding tree in Fig.\,\ref{fig:example} (a), each row shows states with the same number of secondary bonds, i.e., the same potential energy. Two states are connected in the tree if one can topologically transform into the other by making or breaking a single bond. Designing folding protocols, or the order of secondary droplet interactions, allows us to funnel the landscape to one final folded state.

The example of an alternating heptamer chain in Fig.\,\ref{fig:example} (a) shows that switching on only the blue-blue interaction yields a rocket foldamer as the final state. This tree was constructed theoretically  and then populated by images of states that were observed along experimental folding pathways (Methods). The remarkable overlap between experiment and theory indicates that the experiments are sampling all the available states. Tracking $n=255$ folding heptamers allows us to plot the evolution of the yield of the most popular states in each level of the tree in the side panels. Long-lived states correspond to local minima (states $S1$ and $S2$ in the tree) that are theoretical dead-ends, but are overcome in experiments because our system is quasi-two-dimensional and rare out-of-plane rearrangements are possible. As a result, all pathways lead to the rocket foldamer out of the four possible heptamer geometries on a timescale of $\sim 20$ minutes.

\begin{figure*}[t]
  \centering
  \includegraphics{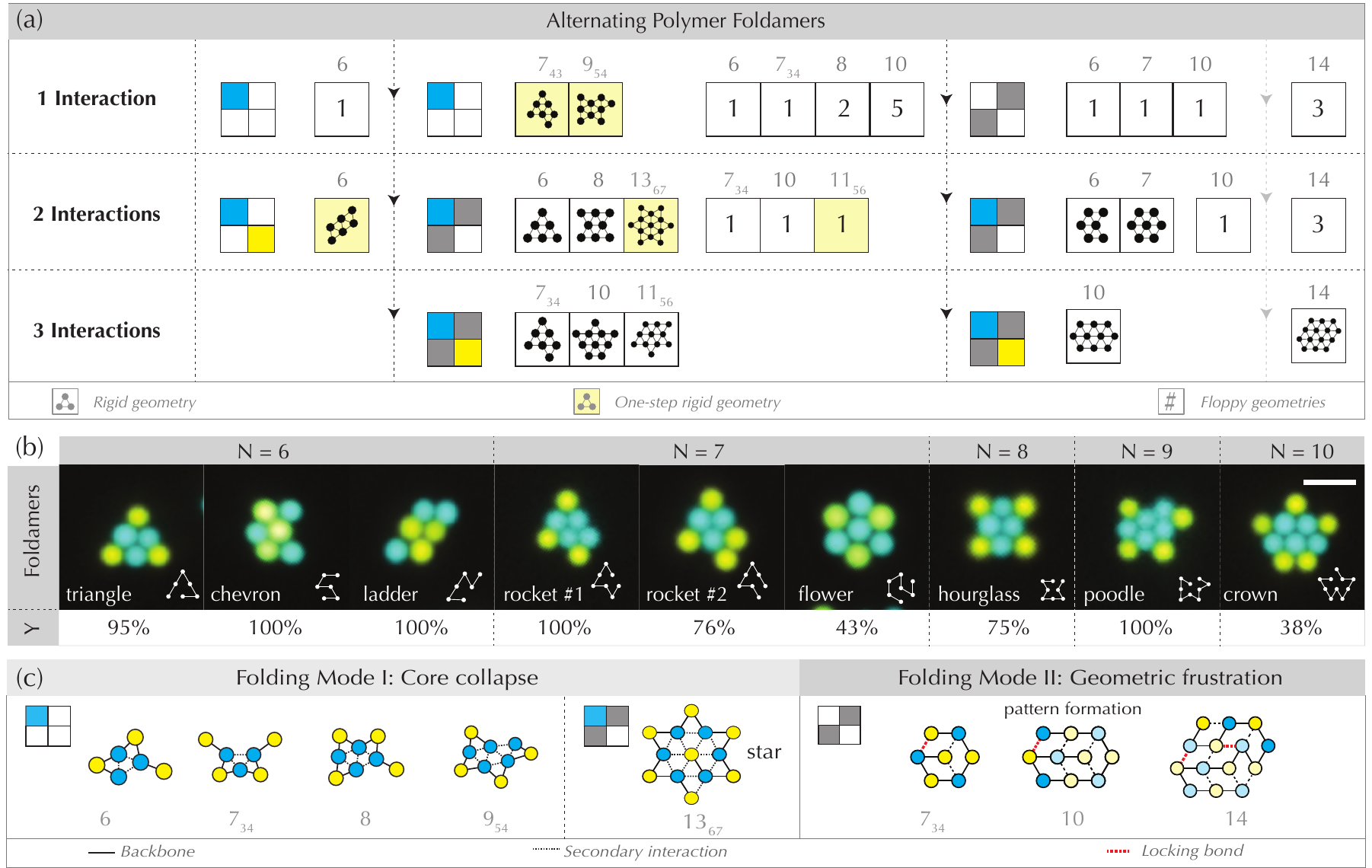}
  \caption{(a) Alternating polymers of length $N=6-14$ (subscripts indicate the number of blue and yellow droplets), can be successfully folded by distinct protocols (columns) with a maximum of three interactions (rows). Foldamers shaded in yellow require only one step, which can switch on one or more interactions. At the end of each step, foldamers are shown on the left and the number of floppy geometries on the right, in order of increasing chain length.  (b) Experimental results show fluorescent images of predicted foldamers up to decamers, as well as their relative folding yields.
  Scale bar is 5\,$\mu$m. (c) Modes of folding: core collapse (left) and geometric frustration (right).}
  \label{fig:protocol}
\end{figure*}

Because the heptamer is comprised of four blue and three yellow droplets, switching on the yellow-yellow interaction funnels the landscape into a much simpler tree, shown in Fig\,\ref{fig:example}\,(b). Here the final state is a unique floppy state that needs additional interactions to become rigid. Subsequently turning on the blue-blue interaction yields two new floppy states, one of which closes into a rigid ladder, while the other requires the remaining blue-yellow interaction to fold into the rocket shape. This particular protocol yields a mixture of the ladder and the rocket and does not qualify as a successful protocol. On the other hand, reversing the order of the last two steps leads only to the rocket foldamer, but with a different color arrangement, or fold, to the one obtained from a single blue-blue interaction in Fig.\,\ref{fig:example} (a). This feature demonstrates the robustness of geometry to the protocol.

\vspace{1mm}
\noindent {\bf Foldamer search algorithm}
\vspace{1mm}

In search of foldamers, we sweep all protocols for folding alternating sequences. The construction of folding trees becomes computationally expensive as the chain length grows, so we devise an alternative strategy for a systematic search that allows us to reach chains with $N \geq 13$ droplets, as shown in the Extended Data Fig.~1. We start by enumerating only the rigid states \cite{Meng2010} and we map out all the possible backbone arrangements therein (Methods). Superimposing the alternating sequence on the backbones, we add secondary bonds between neighboring droplets according to a specific interaction matrix. Resulting states are then classified as local or global minima. Keeping track of the minima each time an interaction is added, we determine if a colloidomer eventually folds into a unique geometry for a given sequence of interactions steps. The algorithm relies on the assumption that interactions are irreversible and that all bonds form, which requires a long enough waiting time at each temperature step in the experiment. This strategy is general for any linear polymer that can freely rearrange during folding via hierarchical interactions.

\vspace{1mm}
\noindent {\bf Alternating sequence foldamers}
\vspace{1mm}

Our systematic search identifies successful protocols that yield a total of eleven foldamer geometries for chains up to 13 droplets long, as shown in Fig.\,\ref{fig:protocol} (a).  Following those protocols, experiments capture most of the predicted foldamers, as shown in Fig.\,\ref{fig:protocol} (b) and Supplementary Videos 1-7. The high relative yields, defined as the proportion of final states that reach the correct geometry, are in excellent agreement with theoretical predictions, as shown in the Extended Data Fig.~2. Exceptions to the high yield are the flower and the crown foldamers, owing to kinetic dead-ends they encounter on timescales beyond the experimental window.

Our foldamers demonstrate that the simplest alternating sequence encodes all the possible geometries of the hexamer: the ladder, the chevron, and the triangle, as shown in Fig.\,\ref{fig:protocol} (b). Among longer foldamers, only the heptamer flower corresponds to the ground state of a folded homocolloidomer, while the rest are unlikely geometries in equilibrium \cite{Trubiano2021}. For example, the octamer hourglass has the highest free energy, i.e., the smallest yield among the nine possible geometries due to its high symmetry number  \cite{Meng2010,Perry2015}. Therefore, our foldamers correspond to kinetic states that are accessible based on geometric considerations alone. Another example is the nonamer poodle, which is the longest chain that can be folded with a single interaction. By contrast, the decamer folds into the crown through a many-to-one transition, as an example of a funnel-like landscape \cite{Bryngelson1995}. 

\newpage
\vspace{1mm}
\noindent {\bf Colloidomer folding mechanisms}
\vspace{1mm}

More generally, alternating colloidomers follow two mechanisms to reach the foldamer state: core collapse and geometric frustration, as illustrated in Fig.\,\ref{fig:protocol} (c). The most common mechanism is the core collapse, which first forms a rigid core and then locks in the remaining droplets on the outside. Up to decamers, the cores consist of a maximum of five identical droplets in unique geometries. Beyond this length, foldamers are comprised of multiflavored cores formed upon turning on two interactions simultaneously, as seen in the star foldamer. 

The second mechanism of geometric frustration initially engages an interaction that traps the droplets via certain locking bonds into positions in which they are surrounded by neighbors with whom they cannot form secondary bonds. Turning on other interactions adds the remaining bonds without changing the geometry. The Russian doll architecture of these foldamers as a function of $N$ allows us to successfully predict the $N=14$ foldamer following the same protocol, shown in Fig. 3(c).

 \begin{figure}
   \centering
   \includegraphics{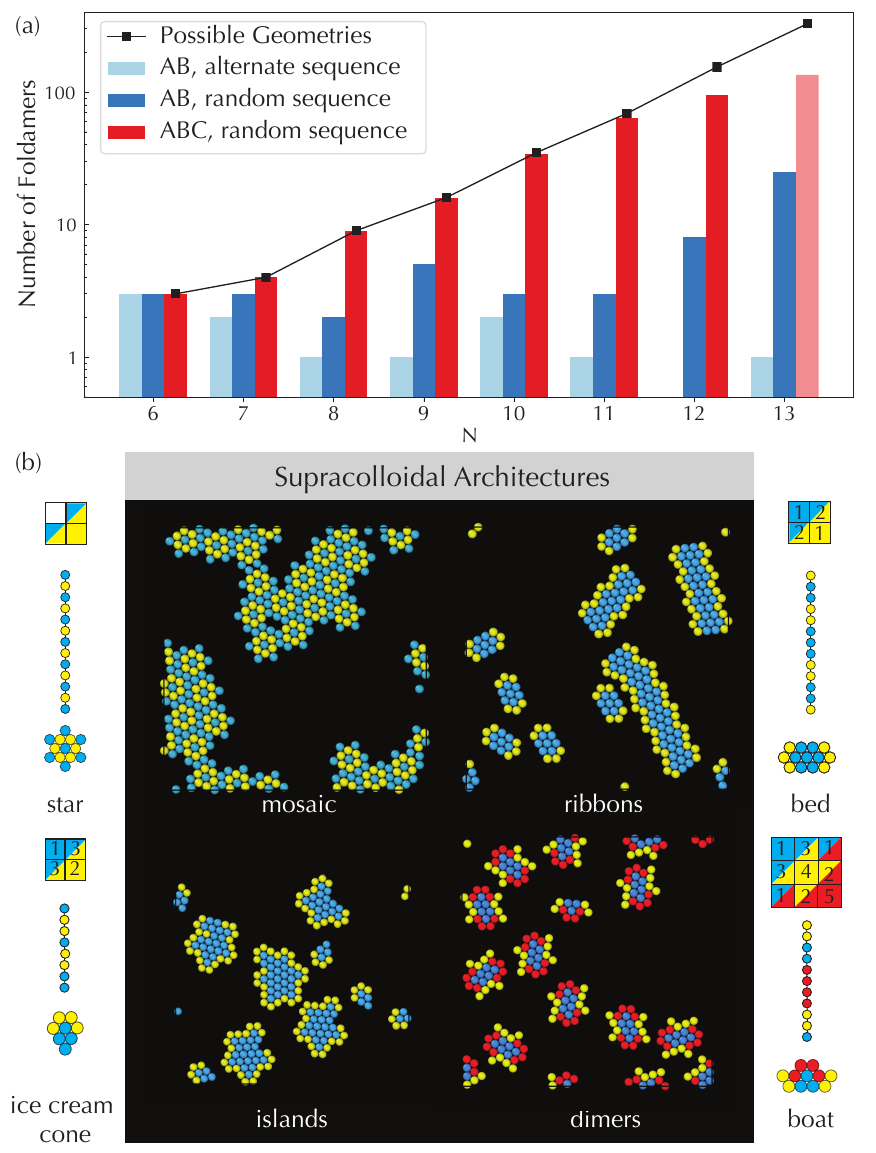} 
 \caption{(a) Exponential growth of the number of possible rigid geometries as a function of chain length  $N$ (black line). Numbers of foldamers encoded by an alternating $AB$ sequence (light blue), any $AB$ sequence (dark blue) and any $ABC$ sequence (red) via all available protocols are shown as bars ($N = 13$ bar is a lower bound).  (b) Simulated examples of supracolloidal architectures self-assembled using specific foldamer interactions. Each foldamer is shown with the protocol and sequence that leads to it. The numbers within the interaction matrix indicate the order of activation of interactions.}
   \label{fig:generalization}
 \end{figure}

\vspace{2mm}
\noindent {\bf From sequence design to large-scale assembly}

Next, we investigate how increasing complexity~\cite{du1998models} improves the number and variety of possible foldamers. We first vary the sequence of droplets while preserving the number of each flavor in the chain. This process uncovers winning protocols that increase the total number of foldamers by roughly an order of magnitude, particularly in longer chains, as shown in Fig.\,\ref{fig:generalization} (a, dark blue). In addition, the introduction of a third flavor while designing both sequence and protocol spaces controls the folding into more than a half of all possible geometries up to tridecamers, totalling $310$ foldamers (red). While two flavors code for all three geometries in hexamers, three letters encode all geometries up to decamers, putting a bound on what can be achieved with a small number of flavors as a function of N~\cite{guarnera2009does, fink2001many}.

With this lexicon of foldamers, we go a step further and use them as building blocks that self-assemble via additional supracolloidal interactions into higher order architectures \cite{Evans2017}, as shown in the simulated examples in Fig\,\ref{fig:generalization} (b). For instance, an interaction between blue droplets assembles star foldamers into a complex mosaic. Foldamers with polarized flavors self-assemble into tubules or colloidal micelles, while three flavors facilitate the assembly of unique dimers. We expect that all these examples are experimentally realizable with further design of supracolloidal interactions.

Our minimal model system exhibits many of the phenomena nominally associated with protein folding. Foldamers comprised of droplets with two or three flavors have the properties of uniqueness \cite{Anfinsen1973}, robustness \cite{Eichler2020}, and kinetic accessibility in a funnel landscape \cite{Bryngelson1995}. The core collapse folding mechanism resembles the hydrophobic collapse in proteins \cite{agashe1995initial, gutin1995burst}, while that of geometric frustration has been proposed as a design principle in the assembly of peptides~\cite{Jiang2017}. On the supracolloidal scale, foldamer assembly mimics the polymerization of fibrils~\cite{Mohapatra2015}, the formation of protein-based micelles~\cite{friedrich2014supramolecular} or protein dimerization~\cite{Marianayagam2004}. These similarities occur even though our system is strictly out-of-equilibrium, highlighting the importance of geometry in guiding assembly.

This bottom up approach gives access to the underlying rules that govern successful folding by dissecting the respective roles of sequence design, minimal number of flavors, hierarchy of interactions, and topological constraints. Moreover, 2D foldamers offer a direct route to folding 3D architectures. Instead of using droplets, one can imagine folding molecular polymers designed with hydrophobic and polar moieties \cite{Hill:2001ic}, 
or building macroscopic beads-on-a-string models with specific interactions, facilitated by an external drive \cite{reches2009folding,Raviv2014}. This new paradigm of hierarchical folding as a precursor for large-scale self-assembly offers design rules for biomimetic materials with tunable functionalities.     
\vspace*{5mm}
\bibliographystyle{apsrev4-1}
\bibliography{bibliography_v1.bib}

\pagebreak
\newpage

\vspace{2mm}
\noindent {\bf Methods} 
\vspace{2mm}
\small

 {\bf Droplet synthesis.} 
Monodisperse PDMS droplets were synthesized according to a protocol modified from that outlined in \cite{McMullen2018}, \cite{McMullen2021}, and \cite{Zhang2017}. 
An equal volume of dimethoxydimethysilane  (Sigma Aldrich) and (3,3,3-trifluoropropyl)methyldimethoxysilane (Gelest) was mixed together with DI water at approximately 2\% v/v. The monomers were prehydrolyzed by vortexing for 60 minutes. 
Ammonia was added at 1\% v/v, and the droplets were left to grow over 24 hours. 
The droplets were then dialyzed against 5mM sodium dodecyl sulfate (SDS, Sigma Aldrich) to remove the remaining ammonia and reaction byproducts.
We then incubated the droplets in 1\% v of \lowercase{(3-GLYCIDOXYPROPYL) METHYLDIETHOXYSILANE} (Gelest) with 10mM Sodium Azide and 5mM SDS. 
This embedded reactive azide groups inside the droplets, such that they can be fluorescently labeled at a later stage. 
This synthesis produced monodisperse oil droplets that were denser than water with a low gravitational height, forming a quasi- 2D system. 

\vspace{2mm}

\textbf{DNA sequences and their interactions.}
The following is a complete list of DNA sequences used in this work, listed with their modifications from 5$^\prime$ to 3$^\prime$.
The strands which formed the interactions were as follows:\\
\noindent \textbf{A:} Azide Cy3A GCA TTA CTT TCC GTC CCG AGA GAC CTA ACT GAC ACG CTT CCC ATC GCT A  GA GTT CAC AAG AGT TCA CAA \\
\noindent \textbf{B:} Azide Cy5 A GCA TTA CTT TCC GTC CCG AGA GAC CTA ACT GAC ACG CTT CCC ATC GCT A  TT GTG AAC TCT TGT GAA CTC\\
\noindent \textbf{C:} Azide AG CAT TAC TTT CCG TCC CGA GAG ACC TAA CTG ACA CGC TTC CCA TCG CTA TTT TTA GTC\\
\noindent \textbf{D:} Azide AG CAT TAC TTT CCG TCC CGA GAG ACC TAA CTG ACA CGC TTC CCA TCG CTA TTT GAC TAA\\
\noindent \textbf{P:} Azide AG CAT TAC TTT CCG TCC CGA GAG ACC TAA CTG ACA CGC TTC CCA TCG CTA TTT ATC GAT\\
\noindent \textbf{CS:} TAG CGA TGG GAA GCG TGT CAG TTA GGT CTC TCG GGA CGG AAA GTA ATG CT Azide

The A and B strands were responsible for the backbone formation and have 20 base long sticky ends. 
In typical experimental conditions, bonds formed by A and B complexation melted at around 75$^\circ$ C.
The C and D strands made a weak complementary interaction, which melted between 30$^\circ$ and 35$^\circ$ C.
This interaction made the $AB$ secondary interaction. 
The P strand formed palindromic self interactions. 
In typical experimental conditions, it melted between 40$^\circ$ and 45 $^\circ$ C. 
The P strand is what gave $AA$ secondary interactions.
Finally, the D strand also had a weak palindromic self interaction. 
In typical experimental conditions, it melted around 27$^\circ$ and provided the $BB$ interaction. 

\vspace{2mm}

{\bf DNA-labeling of emulsion droplets.}
Before labeling with DNA, emulsion droplets were diluted into 1 mM SDS at a volume fraction of approximately 6\%.
DNA strands with sticky ends were reacted with a DBCO terminated pegylated lipid (DPSE-PEG-DBCO, Avanti Polar Lipids), and then annealed with a complementary spacer strand as described in refs \cite{McMullen2018, McMullen2021}. 
Droplets were incubated with backbone DNA at 200 nM concentrations with a volume fraction of 0.6\% with 50 mM NaCl, 10 mM Tris pH 8, and 1 mM EDTA. 
After 30 minutes, secondary interaction DNA was added, bringing the total concentration to 5-25 $\mu M$. 
The droplets were then incubated for two hours before being diluted by a factor of two with a buffer containing 50 mM NaCl, 10 mM Tris pH 8, 0.1\% w/v Triton 165, and Cyanine 3 DBCO (or Cyanine 5 DBCO, both from Lumiprobe). 
The droplets were incubated for a further 30 minutes before being washed several times in 50 mM NaCl to remove all unreacted dye. 

\vspace{2mm}

{\bf Colloidomer formation.}
Droplet polymerization was accelerated by dispersing the droplets in an aqueous ferrofluid (EMG 707, FerroTec) and aligning them with a magnetic field. The ferrofluid was washed several times into 0.3\% F68 pluronic surfactant via centrifugation to remove the proprietary surfactant in the ferrofluid. 
Two sets of droplets were prepared with complementary backbone DNAs and secondary DNA strands of choice. 
The two droplet types were mixed at a 1:1 ratio along with a 1/3 dilution of the F68 ferrofluid buffer, 200 mM NaCl, and 20 mM EDTA pH 8. The sample was added to a custom flow chamber made from a hexamethyldisilazane (Sigma Aldrich) treated glass slide and coverslip, and parafilm. 
The flow cell was sealed with UV glue.

The sample was then heated up to $75^{\circ}$ C to break all bonds in the system, and then cooled down to just above the melting temperature of the strongest secondary interaction, typically $50^{\circ}$ C. 
The sample was then put through a repeated cycle of alignment with rare earth magnets and relaxation in order to grow the chains. 
Typically, this produced a mixed sample of monomers, linear chains, and branched chains. The density of droplets was optimized such that they would grow sizable polymer chains, but that the chains would not aggregate on the timescale of the folding experiments. 
The colloidomers were allowed to relax in the absence of a magnetic field before the folding data was taken. 
Data was taken using a Nikon TI Eclipse with a 20x objective using either single or double channel fluorescence imaging. 

\vspace{2mm}

{\bf Temperature protocols and waiting times.}
The temperature was adjusted using a custom made heating cell composed of a indium tin oxide coated glass slide (SPI) connected to a Thorslabs TC200 resistive heater with a thermocouple for feedback. 
The temperature protocol was programmed through custom software. 
For a given temperature protocol, first a sample of droplet polymers with the desired set of interactions was made.
A manual sweep of the temperature was performed to determine where each interaction takes place, since the melting temperatures can change from sample to sample. 
The first temperature step lasting 10 minutes was programmed to be above the melting temperature of all interactions to identify the unfolded colloidomers. 

Subsequently, there can be one, two, or three additional steps depending on how many interactions are to be turned on. 
If there is more than one interaction that is turned on, the waiting step for the first interaction is the longest. For the data in Fig. 3 (c), the waiting time at the first step was 20 minutes (except for the N = 6 triangle, which had a waiting time of 30 minutes), while that for the second and third steps was typically 5 to 10 minutes.
In principle, longer waiting times allow for the resolution of local minima and lead to better yields. 
In practice, however, longer waiting times increase the chance that colloidomers aggregate during folding, which can be avoided in dilute samples. 

\vspace{4mm}

{\bf Video analysis.}
Folding videos were analyzed using a custom MATLAB data analysis software. All particles were identified and located using thresholding. These particles were then tracked through the whole movie using custom software modeled after that in \cite{guerra2018freezing}. Polymers were identified using the same metrics as in~\cite{McMullen2018} from the first ten minutes of every recording, which was always above the melting temperature of the strongest secondary interaction. 
A $N \times N \times t$ (where $N$ is the number of monomers in the polymer and $t$ is the time) connectivity matrix was then calculated for each polymer using the particle locations and diameters. 
The contact matrix was median filtered over $t$ to remove transient interactions. 
Each contact matrix was then matched to a polymer configuration theoretically computed, allowing us to track the polymer configuration over time. Selections of data were vetted by hand afterwards to ensure the integrity of the data. 
Polymers that aggregated or that folded into three-dimensional structures were discarded. 

For Figure 2 (a), the yield plotted is defined as the proportion of all polymers that are identified with a given configuration. If a polymer is lost at a given time, aggregates with another one, or enters an unidentifiable configuration, it is removed from the pool. 
For Figure 3 (c), the yield is defined as the fraction of polymers of length $N$ that fold to completion into the target structure over the fraction of polymers of length $N$ that fold to completion into any structure.

\vspace{2mm}

\textbf{Enumerating two-dimensional geometries.}
We define as a geometry any colloidomer cluster where deformations cost energy, i.e., a deformation requires breaking a secondary bond. Geometries are therefore rigid clusters. To enumerate two-dimensional geometries for a system of size $N$, we start by selecting all possible sets of $N$ neighbouring points on a $N\times N$ triangular lattice. We form bonds between points located at a unit distance and test the rigidity of resulting geometries by analyzing the normal modes of the dynamical matrix. We describe the ensemble of geometries for a chain of length $N$ by a set of planar graphs $\{G_{i, N}(V, E)\}$, $i \in (1, N_{R})$, where the vertices are the droplets in the chain and the edges are the DNA-mediated bonds. Edges may be of two types: backbone bonds and secondary bonds. Each graph is characterized by a contact matrix, which describes the bonds between droplets, and a distance matrix, which contains the distances between each droplet pair in a geometry. The first size with more than one geometry is $N = 6$ \cite{Perry2015}. At $N\geq 13$ the first geometries with stable holes in the bulk appear. 

\vspace{2mm}

\textbf{Foldamer search algorithm.}
We develop a computationally efficient search algorithm to systematically scan protocol and sequence spaces and find foldamers of a given length $N$. The algorithm requires as input the ensemble of all backbone configurations within the geometries $N_R$ for a chain of length $N$, i.e., the set of Hamiltonian paths $\{ H_{1, 1}, ...,  H_{p_1, 1}, ..., H_{1, q}, ..., H_{p_q, q}\}$, $\forall q \in (1, N_R)$, where $p_q$ is the number of paths in the $q$-th geometry. The total number of Hamiltonian paths grows exponentially and it does not depend on the sequence or the interaction matrix. Thus, they are computed only once per $N$,  significantly reducing the computation time. The structure of the algorithm is shown in the Extended Data Fig.~1.  
For a given protocol and sequence, the algorithm can be summarized as follows:
\vspace{2mm}

\noindent {\bf Input.} {Map the sequence onto Hamiltonian paths.} \\
\noindent {\bf 1.} Form bonds. Apply the first interaction of the protocol. A bond will be formed between two vertices if they are in neighbouring lattice points and the interaction is allowed\\
\noindent {\bf 2.} {Are there geometries?}
    \begin{enumerate}
        \item[{\bf (a)}] Yes. If the classification flags geometries, the algorithm stops. If there is a single geometry, a foldamer is reported. We choose to report a solution even if there are competing floppy states with the same or more bonds as the foldamer geometry (this becomes possible when $N\ge7$). 
        \item [{\bf (b)}]{No. A foldamer is not selected.}
        \begin{enumerate}
            \item[{\bf 3.}] {Select global minima}. This is analogous to selecting floppy states with the largest number of bonds. Note that this also implies that local minima in the first interaction tree are not considered (we assume here strict downhill folding).  
            \item[{\bf 4.}] Continue the protocol of adding interactions. {Update the interaction matrix according to the protocol.}
            \item[{\bf 5.}] Form new bonds. {Repeat the bond-making process iterating over the states from step 3.} 
            \item[{\bf 6.}] {Classify states.} We classify states into global and local minima, and transient states. Global minima are states of a tree that cannot acquire additional bonds either because they reached a rigid state, or because spatially accessible neighbors do not have flavors with attractive interactions. Local minima are floppy states whose topology prevents further formation of bonds. All other states are classified as transient states.
            \item[{\bf 7.}] Is the protocol over? 
            \begin{enumerate}
            \item[i)] Yes. Analyze the resulting geometries. If a single geometry is found, a foldamer is reported.
            \item[{ii)}] No. {Repeat steps 4-7 until the protocol ends.}
            \end{enumerate}
            \end{enumerate}
    \end{enumerate}

\vspace{2mm}

\textbf{Simulation details.} We perform Dissipative Particle Dynamics (DPD) \cite{Groot} simulations using an in-house code. Our unit of length is the particle diameter $\sigma = 1$ and we assume all particles have the same mass $m = 1$. Energy is measured in units of $k_BT$ and we fix the temperature of the system at $k_BT = 1$.  When folding a colloidomer of length $N$, we set the simulation box size to $L/\sigma = (N + 2) $. For the self-assembly of supracolloidal architectures, we choose $L/\sigma = 30$. In both cases we use periodic boundary conditions. We use a multiple-step simulation scheme to integrate the equations of motion with $dt_{s} = 10^{-2}$ to resolve the dynamics of the solvent and $dt_{c} = 10^{-4}$ for the dynamics of the colloids. DNA-mediated interactions are modelled via a short-range, isotropic interaction potential \cite{Wang}
\begin{equation}
U(r) = \varepsilon \alpha(r_i, \sigma) \left[ \left( \frac{\sigma}{r} \right)^2 -1 \right]\left[ \left( \frac{r_i}{r} \right)^2 -1 \right]^ 2,
\end{equation}
where $r_i = 1.05\sigma$ is the interaction range, $\varepsilon$ is the strength of the interaction and $\alpha$ is a parameter that sets the minimum of the potential $U(r_{min}) = \varepsilon$ (see \cite{Wang} for further details). Primary bonds are made irreversible by setting $\varepsilon_{P} = 40 k_B T$. To simulate secondary interactions, we gradually increase $\varepsilon_{S}$ until it reaches $\varepsilon_B$, once the corresponding interaction is turned on. The increase is done over the course of 200 simulation steps to ensure downhill folding while preventing poor potential sampling.  

{\bf Acknowledgments:} 
 The authors would like to thank David Pine, Alexander Grosberg, Paul Chaikin, Sascha Hilgenfeldt, Eric Cl\'ement, Olivier Rivoire and Ludwik Leibler. This work was supported by the Paris Region (Région Île-de-France) under the Blaise Pascal International Chairs of Excellence. This work was also supported by the MRSEC program of the National Science Foundation under Grants No. NSF DMR-1420073, No. NSF PHY17-48958, and No. NSF DMR-1710163, as well as the European Union’s Horizon 2020 research and innovation program under the Marie Skłodowska-Curie grant agreement No. 754387.

{\bf Author Contributions:}  
A.M. designed the materials, synthesized the droplet system, performed the folding experiments and developed data analysis tools to extract time-dependent yields of foldamers. M.M.B. and Z.Z. constructed the theoretical model for generating folding trees and rigid states. M.M.B. developed the algorithm for enumerating foldamers, wrote the molecular dynamics codes, and performed numerical simulations. J.B. and Z.Z. conceived the study and supervised the research. The manuscript was written by J.B. and Z.Z. together with A.M and M.M.B.

{\bf Competing Interests:} 
 The authors declare no competing interests. 
 
 {\bf Data Availability :} 
 The data that support the findings of this study are available from the corresponding author upon reasonable request. The data includes experimental as well as computational datasets, Matlab scripts for experimental video analysis and Python scripts for computational dataset analysis.
 
 {\bf Code Availability:}
 The custom computer codes to build folding trees, to identify foldamers and the Dissipative Particle Dynamics code to simulate folding of colloidomers are available from from the corresponding author upon reasonable request.

{\bf Supplementary Information:} 
Supplementary Information is available for this paper.

{\bf Correspondence and requests for materials} should be addressed to Jasna Brujic or Zorana Zeravcic.

\clearpage
\newpage
\renewcommand{\figurename}{{\bf Extended Data Fig.}}
\setcounter{figure}{0}  

\newpage

\begin{figure}[htp!]
    \centering
    \includegraphics[width=0.49\textwidth]{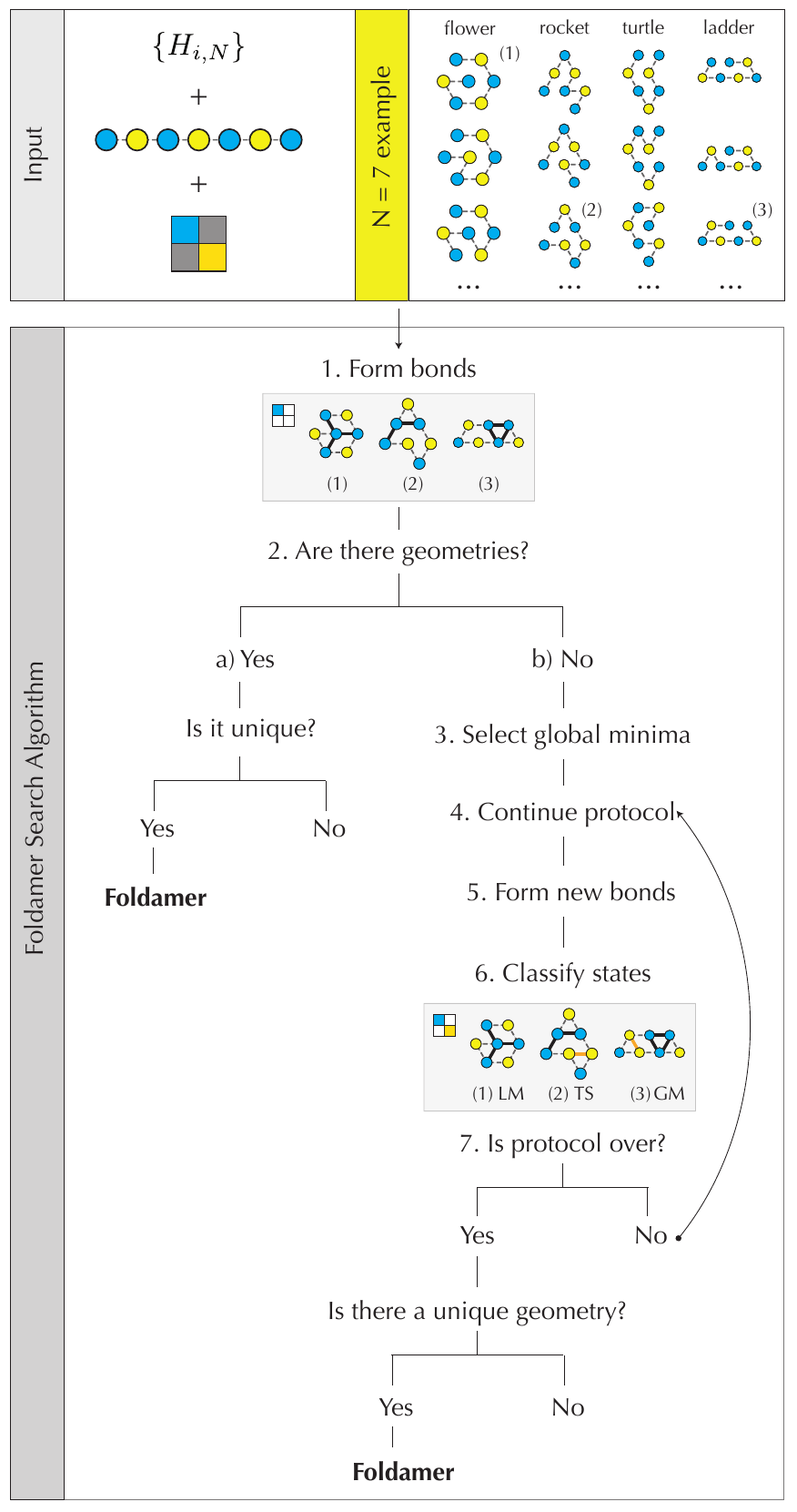}
    \caption{{\bf Flowchart of the foldamer search algorithm.} The top panel shows the ingredients required to run the algorithm: the ensemble of Hamiltonian paths $\{ H_{i,N}\}$, a sequence, and a protocol. The input is a set of colored Hamiltonian paths embedded on the geometries, as shown for $N=7$ and an alternate $ABABABA$ sequence. The bottom panel outlines the main steps of the algorithm.}
    \label{fig:alg}
\end{figure}

\newpage

\begin{figure*}[htp!]
    \centering
    \includegraphics[width=\textwidth]{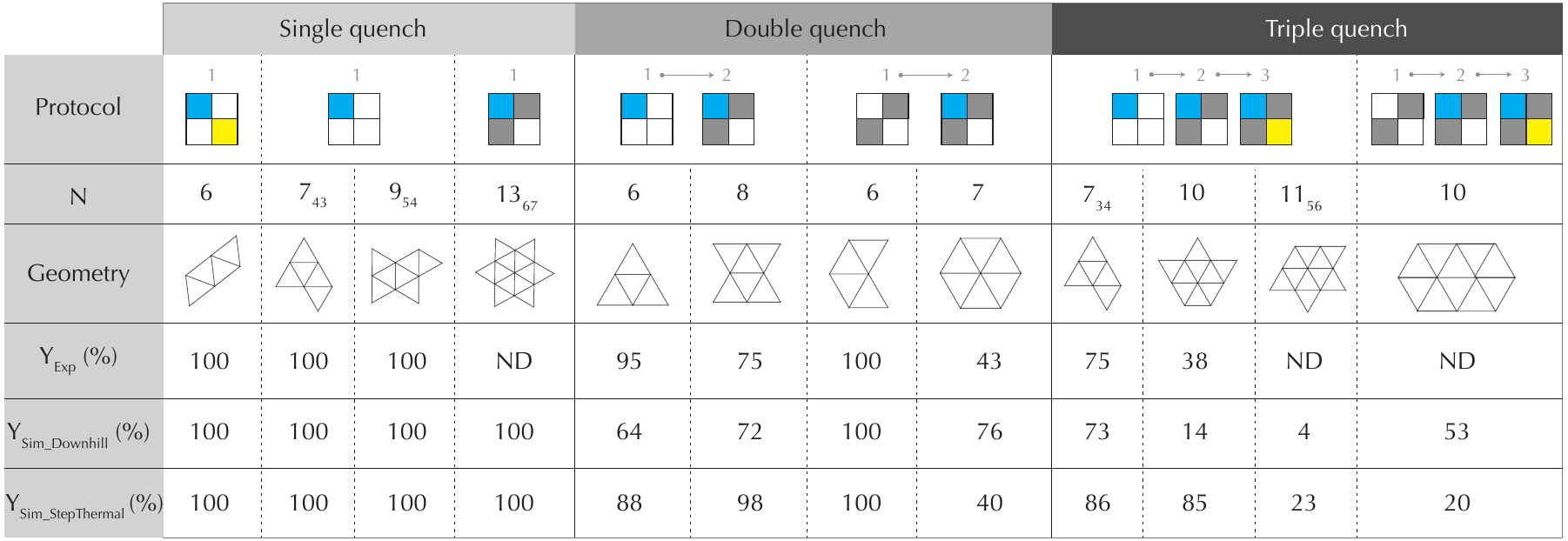}
    \caption{{\bf Foldamer yields for an alternating $ABAB$ sequence with length $N = 6-13$.} From left to right, we show the results for single, two, and three-step protocols. All yields are given as relative yields, in which the number of foldamers is normalized by the total number of rigid structures observed at the end of the corresponding protocol. The experimental number of observations is $n_6 \text{ }[\text{ladder}, \text{triangle}, \text{chevron}] = (67, 19, 86)$, $n_7 \text{ }[\text{rocket \#1}, \text{rocket \#2}, \text{flower}] = (175, 25, 7) $, $n_8 \text{ }[\text{hourglass}] = 8$, $n_9 \text{ }[\text{poodle}] = 24$ and $n_{10} \text{ }[\text{crown}] = 8$. `ND' stands for `No Data'. These experimental data are in good agreement with numerical simulations. Purely downhill simulations optimize the yield $Y_{Sim\_Downhill}$ of geometrically frustrated foldamers, such as the flower and the bed. Repeating the simulations on timescales where some rearrangements are possible optimizes the yield $Y_{Sim\_StepThermal}$ of core collapse foldamers, as shown in Supplementary Video 8. For simulations with multistep protocols, the waiting time between subsequent interactions is $\tau = 10^5$ simulation time units. The total number of simulations is $>2\times 10^3$ for all cases reported.}
    \label{fig:tableyields}
\end{figure*}

\end{document}